\documentstyle[amssymb,aps]{revtex}

\input epsf

\begin{document}
\title{Anomalous statistical properties of the critical current distribution in
superconductor containing fractal clusters of a normal phase}
\author{Yuriy I. Kuzmin}
\address{Ioffe Physical Technical Institute of the Russian Academy of Sciences,\\
Polytechnicheskaya 26 St., Saint Petersburg 194021 Russia,\\
and State Electrotechnical University of Saint Petersburg,\\
Professor Popov 5 St., Saint Petersburg 197376 Russia\\
e-mail: yurk@mail.ioffe.ru; iourk@yandex.ru\\
tel.: +7 812 2479902; fax: +7 812 2471017}
\date{\today}
\maketitle
\pacs{74.60.Ge; 74.60.Jg; 05.45.Df; 61.43.Hv}

\begin{abstract}
Statistical properties of the critical current distribution in
superconductor with fractal clusters of a normal phase are considered. It is
found that there is the range of fractal dimensions in which the variance
and expectation for this distribution increases infinitely. Simple technique
of avoiding such a divergence by the use of truncated distributions is
proposed. It is suggested that the most current-carrying capability of a
superconductor can be achieved by modifying the cluster area distribution in
such a way that the regime of giant variance of critical currents will be
realized.
\end{abstract}

\section{INTRODUCTION}

Considerable recent attention is being drawn to the fractal behavior of
magnetic flux in type-II superconductors \cite{surdeanu}-\cite{prb}. High
temperature superconductors (HTS) containing clusters of correlated defects
\cite{bazil,mezzetti} are of special interest in this field. The case of
clusters with fractal boundaries provides new possibilities for increasing
the critical current value \cite{pla,tpl}. By virtue of the capability to
trap a magnetic flux, such clusters can appreciably modify magnetic and
transport properties of superconductors \cite{prb,pla2,pss}. The
distribution of the critical currents in superconductor containing the
normal phase clusters with fractal boundaries has unusual statistical
properties, and it is these features that will interest us.

Let us consider a superconductor containing columnar inclusions of
a normal phase, which are out of contact with one another. These
inclusions may be formed by the fragments of different chemical
compositions, as well as by the domains of the reduced
superconducting order parameter. The similar columnar defects can
readily be created during the film growth process \cite
{mezzetti,pjtf99,ftt99}. In the course of the cooling below the
critical temperature in the magnetic field (``field-cooling``
regime) the magnetic flux will be trapped in the isolated clusters
of a normal phase so the two--dimensional distribution of the flux
will be created in such a superconducting structure. When the
transport current is passed transversely to the magnetic field,
this one is added to all the persistent currents, which circulate
around the normal phase clusters and keep the trapped magnetic
flux to be unchanged. By the cluster we mean a set of the columnar
defects, which are united by the common trapped flux and are
surrounded by the superconducting phase. Inasmuch as the
distribution of the trapped magnetic flux is two--dimensional,
instead of dealing with an extended object, which indeed the
normal phase cluster is, we will consider its cross-section by the
plane carrying a transport current. As was first found in
Ref.~\cite{pla}, clusters of a normal phase can have fractal
boundaries, and this feature has a significant effect on the
dynamics of the trapped magnetic flux \cite{prb,pss}. In the
subsequent consideration we will suppose that the characteristic
sizes of the normal phase clusters far exceed both the coherence
length and the penetration depth. This assumption agrees well with
the data on the cluster structure in YBCO films
\cite{pla,prb,ftt99}, as well as will allow us to highlight the
role played by the cluster boundary in the magnetic flux trapping.

\section{GIANT DISPERSION OF CRITICAL CURRENTS}

Suppose that there is a superconducting percolation cluster in the plane of
the film where a transport current flows. Such a structure provides for an
effective pinning, because the magnetic flux is locked in finite clusters of
a normal phase. When the transport current is increased, the trapped
magnetic flux remains unchanged until the vortices start to break away from
the clusters of pinning force weaker than the Lorentz force created by the
current. As this takes place, the vortices must cross the surrounding
superconducting space, and they will first do that through the weak links,
which connect the normal phase clusters between themselves. Such weak links
form easily in HTS characterized by an extremely short coherence length.
Diverse structural defects, which would simply cause an additional
scattering at long coherence length, give rise to the weak links in HTS.
Weak links arise readily on twin boundaries, and magnetic flux can easily
move along them \cite{duran}. Whatever the microscopic nature of weak links
could be, they form the channels for vortex transport. Accordingly to weak
link configuration each normal phase cluster has its own value of the
critical current, which contributes to the total distribution. By the
critical current of the cluster we mean the current of depinning, that is to
say, such a current at which the magnetic flux ceases to be held inside the
cluster of a normal phase. The critical current distribution is related to
the cluster area distribution, because the cluster of a larger size has more
weak links over its boundary with the surrounding superconducting space, and
thus the smaller current of depinning \cite{prb}.

\begin{figure}[tbp]
\epsfbox{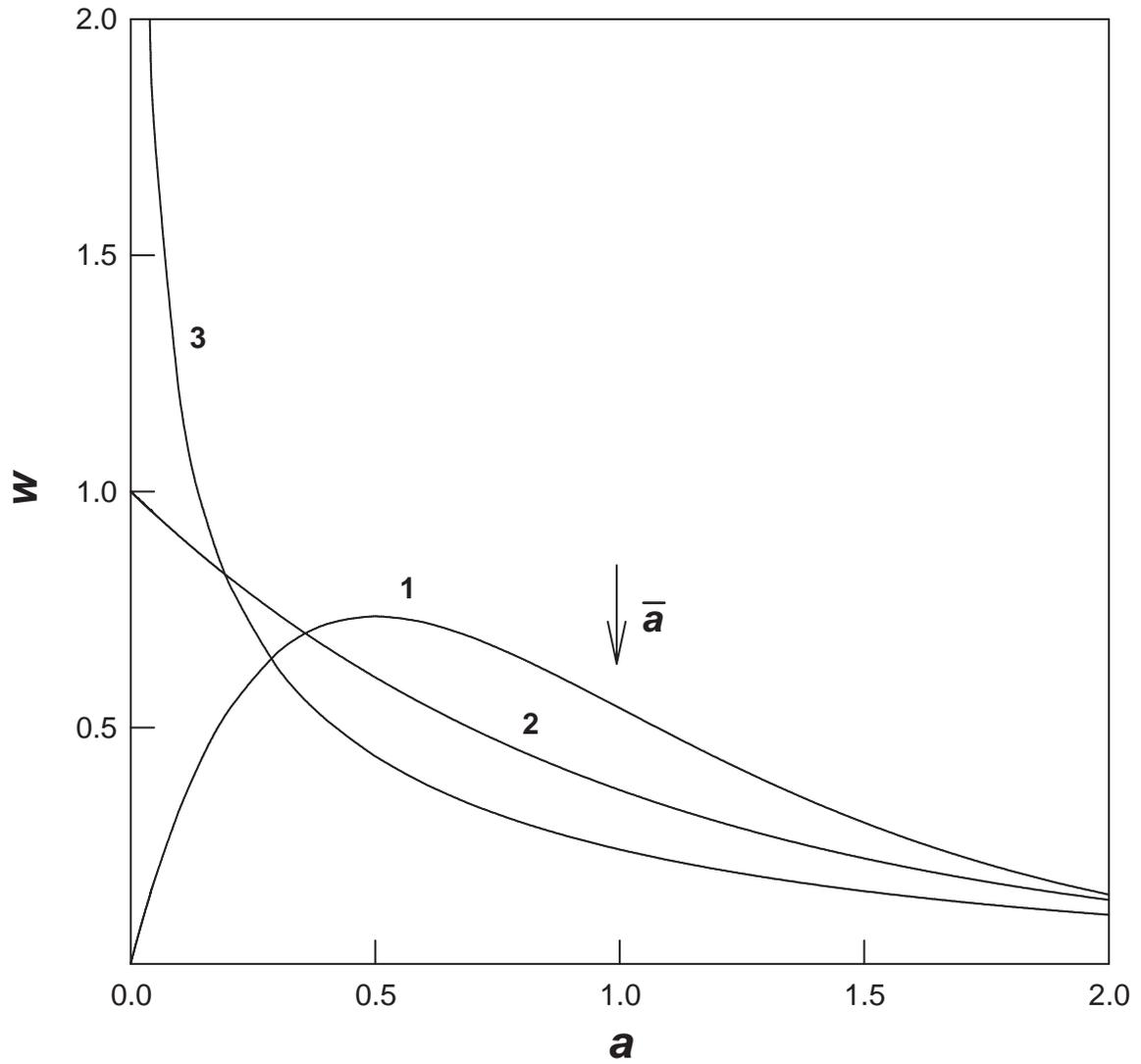}
\caption{The distribution of the areas of normal phase clusters at different
values of $g$-parameter. Curve (1) corresponds to the case of $g=1$;
curve (2) of $g=0$; curve (3) of $g=-0.5$. The arrow indicates the
mean cluster area.}
\label{figure1}
\end{figure}

In the practically important case of YBCO films with columnar defects the
exponential distribution of the cluster areas can be realized \cite{pla},
which is the special case of gamma distribution. The exponential
distribution has only one characteristic parameter (mean cluster size), so
there are not many possibilities to modify the geometric morphological
properties of the clusters in this simplest case. By contrast, in the case
of gamma distribution there is an additional way for optimizing the cluster
structure of the composite superconductor by the control of two independent
parameters in the course of the film growth. One of the aims of the present
work is to find how the cluster area distribution should be optimized in
order to get the highest current-carrying capability of a superconductor. In
Ref.~ \cite{pla2} the critical current distribution was derived in the case
of gamma distribution of fractal clusters areas, which has the following
probability density
\begin{equation}
w\left( A\right) =\frac{A^{g}\exp \left( -A/A_{0}\right) }{\Gamma \left(
g+1\right) A_{0}^{g+1}}  \label{dens1}
\end{equation}
where $\Gamma (\nu )$ is Euler gamma function, $A$ is the cluster area, $%
A_{0}>0$ and $g>-1$ are the parameters of gamma distribution that control
the mean area of the cluster $\overline{A}=(g+1)A_{0}$ and its variance $%
\sigma _{A}^{2}=\left( g+1\right) A_{0}^{2}$. For further consideration it
is convenient to introduce the dimensionless area of the cluster $a\equiv A/%
\overline{A}$, for which the distribution function of Eq.~(\ref{dens1}) can
be rewritten as:
\begin{equation}
w\left( a\right) =\frac{\left( g+1\right) ^{g+1}}{\Gamma \left( g+1\right) }%
a^{g}\exp \left( -\left( g+1\right) a\right)  \label{densless2}
\end{equation}
The mean dimensionless area of the cluster is equal to unity, whereas the
variance is determined by one parameter only: $\sigma _{a}^{2}=1/\left(
g+1\right) $. The probability density of Eq.~(\ref{densless2}) is presented
in Fig.~\ref{figure1} for the characteristic values of $g$-parameter. The
critical current distribution has the following form \cite{pla2}
\begin{equation}
f(i)=\frac{2G^{g+1}}{D\Gamma (g+1)}i^{-\left( 2/D\right) (g+1)-1}\exp \left(
-Gi^{-2/D}\right)  \label{cur3}
\end{equation}

where
\[
G\equiv \left( \frac{\theta ^{\theta }}{\theta ^{g+1}-\left( D/2\right) \exp
\left( \theta \right) \Gamma (g+1,\theta )}\right) ^{\frac{2}{D}}
\]

\[
\theta \equiv g+1+\frac{D}{2}
\]

$\Gamma (\nu ,z)$ is the complementary incomplete gamma function, $i\equiv
I/I_{c}$ is the dimensionless electric current, $I_{c}=\alpha \left(
A_{0}G\right) ^{-D/2}$ is this the critical current of the resistive
transition, $\alpha $ is the form factor, and $D$ is the fractal dimension
of the cluster perimeter. The value of $D$ specifies the scaling relation $%
P^{1/D}\propto A^{1/2}$ between perimeter $P$ and area $A$ of the cluster
\cite{mandelbrot,mandelbrot2}.

\begin{figure}[tbp]
\epsfbox{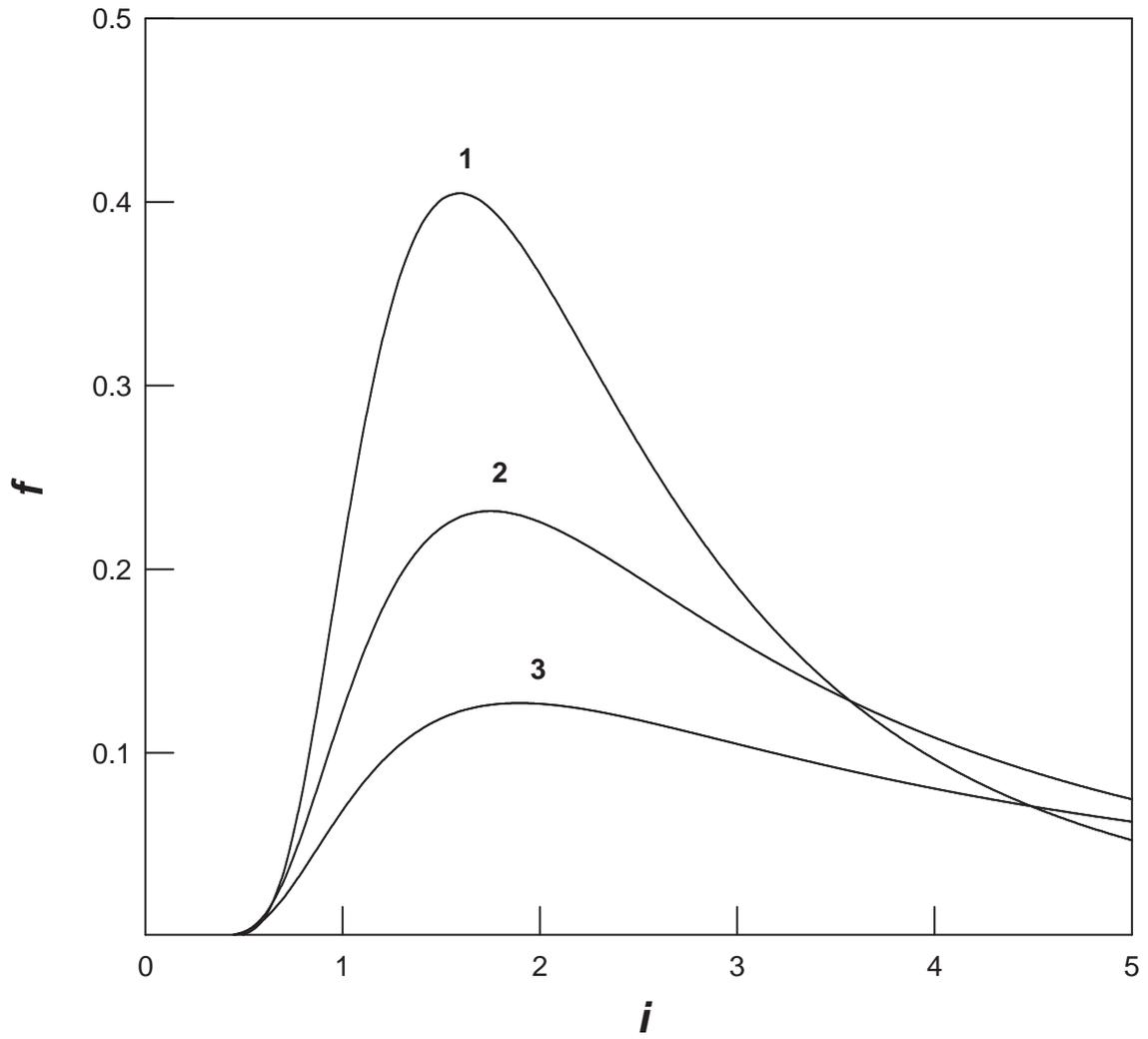}
\caption{The critical current distribution
for the fractal dimension of the cluster boundary $D=1.5$. Curve
(1) corresponds to the case of $g=1$; curve (2) of $g=0$; curve
(3) of $g=-0.5$.}
\label{figure2}
\end{figure}

\begin{figure}[tbp]
\epsfbox{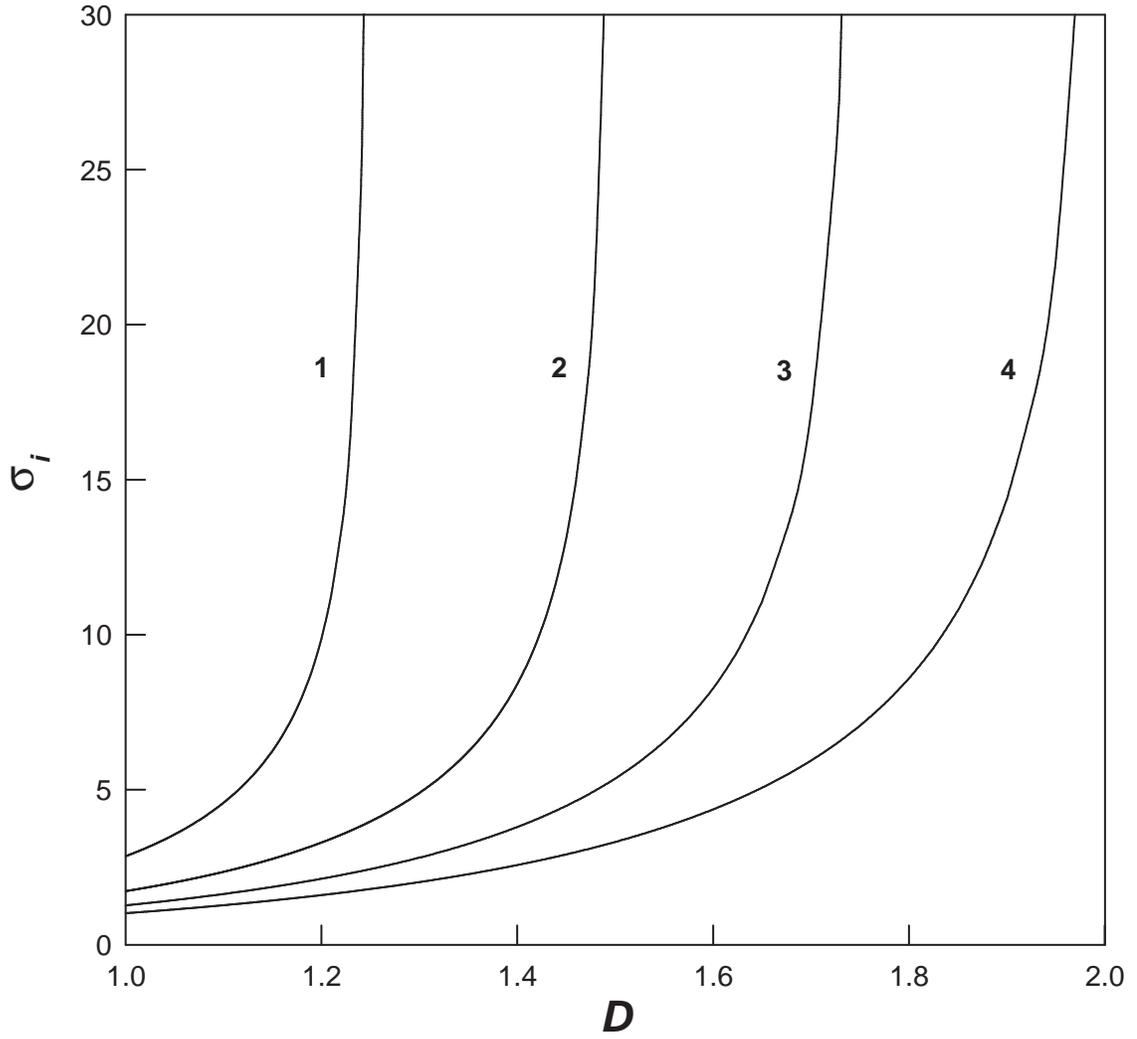}
\caption{Influence of the fractal dimension of the cluster boundary on the
standard deviation of critical currents. Curve (1) corresponds to the case
of $g=0.25$; curve (2) of $g=0.5$; curve (3) of $g=0.75$; curve
(4) of $g=1$.}
\label{figure3}
\end{figure}

The probability density curve for critical current distribution of Eq.~(\ref
{cur3}) has the skew bell-shaped form with inherent broad ``tail`` extended
over the region of high currents (see Fig.~\ref{figure2}). As may be seen
from this graph, the critical current distribution spreads out with some
shift to the right as $g$-parameter decreases. This broadening can be
characterized by the standard deviation of critical currents
\begin{equation}
\sigma _{i}=G^{\frac{D}{2}}\sqrt{\frac{\Gamma \left( g+1-D\right) }{\Gamma
\left( g+1\right) }-\left( \frac{\Gamma \left( g+1-D/2\right) }{\Gamma
\left( g+1\right) }\right) ^{2}}  \label{sd4}
\end{equation}
The standard deviation grows nonlinearly with increase in the fractal
dimension, as is illustrated in Fig.~\ref{figure3}. The peculiarity of the
distribution of Eq.~(\ref{cur3}) is that its variance becomes infinite in
the range of fractal dimensions $D\geq g+1$. The distributions with
divergent variance are known in probability theory - the classic example of
that kind is Cauchy distribution \cite{hudson}. However, such an anomalous
feature of exponential-hyperbolic distribution of Eq.~(\ref{cur3}) is of
special interest, inasmuch as the current-carrying capability of a
superconductor would be expected to increase in the region of giant
variance. Then the statistical distribution of critical currents has a very
elongated ``tail`` containing the contributions from the clusters of the
highest depinning currents.

The distribution of Eq.~(\ref{cur3}) has one even more striking feature: its
mathematical expectation, which represents the mean critical current
\begin{equation}
\overline{i}=\frac{\Gamma \left( g+1-D/2\right) }{\Gamma \left( g+1\right) }%
G^{\frac{D}{2}}  \label{mean5}
\end{equation}
is also divergent in the range of fractal dimensions $D\geqslant 2\left(
g+1\right) $. At the same time, the mode of the distribution {\it mode}$%
\,f(i)=\left( G/\theta \right) ^{D/2}$ remains finite for all possible
values of the fractal dimension $1\leqslant D\leqslant 2$.

The reason for divergence of the mean critical current consists in the
behavior of the cluster area distribution of Eq.~(\ref{densless2}). As may
be seen from Fig.~\ref{figure1}, the graph of the distribution function of
Eq.~(\ref{densless2}) takes essentially different shapes depending on the
sign of $g$-parameter - from the skew unimodal curve (1) (at $g>0$) to the
monotonic curve (3) with hyperbolic singularity at zero point (at $g<0$). In
the borderline case of $g=0$, which separates these different kinds of the
functions, the distribution obeys an exponential law (curve (2)). It is just
for negative values of $g$-parameter that the mean critical current
diverges. The contribution from the clusters of small area to the overall
distribution grows at $g<0$. Since the clusters of small size have the least
number of weak links over a perimeter, they can best trap the magnetic flux.
Therefore, an increase of the part of small clusters in the area
distribution of Eq.~(\ref{densless2}) leads to a growth of the contribution
with high depinning currents made by these clusters in the critical current
distribution of Eq.~(\ref{cur3}). Just as a result of this feature the mean
critical current diverges at $g<0$. Nevertheless, the total area between the
curve of the probability density (like curve (3) in Fig.~\ref{figure2}) and
the abscissa remains finite by virtue of the normalization requirement.

\section{TRUNCATED DISTRIBUTION OF CRITICAL CURRENTS}

Obviously, the proper critical current cannot be infinitely high as well as
the clusters of infinitesimal area do not really exist. There is the minimum
value of the normal phase cluster area $A_{m}$ , which is limited by the
processes of the film growth. So in YBCO based composites, which were
prepared by magnetron sputtering on sapphire substrates with a cerium oxide
buffer sublayer \cite{prb}, the sample value of minimum area of the normal
phase cluster has been equal to $A_{m}=2070~$nm$^{2}$ at mean cluster area $%
\overline{A}=76500~$nm$^{2},$ that corresponds to the lower bound of the
dimensionless area of the cluster $a_{m}\equiv A_{m}/\overline{A}=0.027$. In
view of this limitation, we will describe the distribution of the normal
phase cluster areas by the truncated version of the probability density of
Eq.~(\ref{densless2}):
\begin{equation}
w\left( a\left| a\geqslant a_{m}\right. \right) =\frac{h\left(
a-a_{m}\right) }{1-W\left( a_{m}\right) }w\left( a\right)  \label{truna6}
\end{equation}
where $\gamma \left( \nu ,z\right) $ is the complementary gamma function, $%
h\left( x\right) \equiv \left\{
\begin{array}{cc}
1 & \ \ for\ \ x\geqslant 0 \\
0 & \ \ for\ \ x<0
\end{array}
\right. $ \ \ \ is the Heaviside step function,
\begin{equation}
W\left( a_{m}\right) \equiv \int\limits_{0}^{a_{m}}w\left( a\right) da=\frac{%
\gamma \left( g+1,\left( g+1\right) a_{m}\right) }{\Gamma \left( g+1\right) }
\label{degr7}
\end{equation}
is the truncation degree, which is equal to the probability $\Pr \left\{
\forall a<a_{m}\right\} $ to find the cluster of area smaller than the least
possible value $a_{m}$ in the initial population.

The expression of Eq.~(\ref{truna6}) gives the conditional distribution of
probability, for which all the events of finding a cluster of area less than
$a_{m}$ are excluded. Thus the truncation provides a natural way to fulfil
our initial assumption that the cluster size has to be greater than the
coherence and penetration lengths. New distribution of cluster areas gives
rise to the truncated distribution of the critical currents:
\begin{equation}
f\left( i\left| i\leqslant i_{m}\right. \right) =\frac{h\left(
i_{m}-i\right) }{1-W\left( a_{m}\right) }f\left( i\right)  \label{truni8}
\end{equation}
where $i_{m}\equiv \left( G/\left( g+1\right) a_{m}\right) ^{D/2}$ is the
upper bound of the depinning current, which corresponds to the cluster of
the least possible area $a_{m}$.

Then, instead of Eqs.~(\ref{sd4}) and (\ref{mean5}), the standard deviation
and mean critical current are
\[
\sigma _{i}^{\ast }=G^{\frac{D}{2}}\sqrt{\frac{\Gamma \left( g+1-D,\left(
g+1\right) a_{m}\right) }{\Gamma \left( g+1,\left( g+1\right) a_{m}\right) }%
-\left( \frac{\Gamma \left( g+1-D/2,\left( g+1\right) a_{m}\right)
}{\Gamma \left( g+1,\left( g+1\right) a_{m}\right) }\right) ^{2}}
\]

\[
\overline{i}^{\ast }=\frac{\Gamma \left( g+1-D/2,\left( g+1\right)
a_{m}\right) }{\Gamma \left( g+1,\left( g+1\right) a_{m}\right) }G^{\frac{D}{%
2}}
\]

\begin{figure}[tbp]
\epsfbox{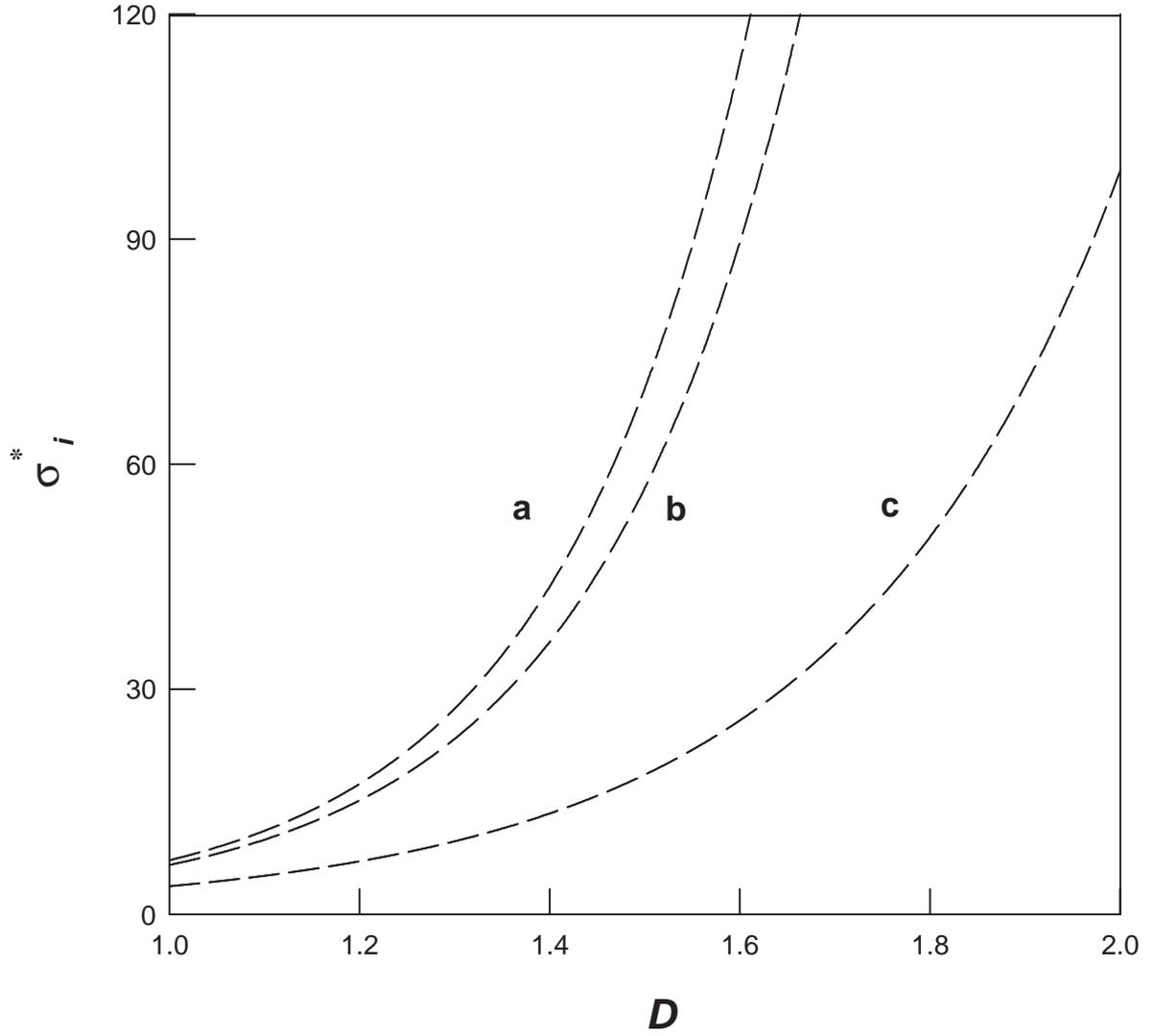}
\caption{Standard deviation graphs for
truncated distribution of the critical currents at $g=-0.2$ with
the different degree of truncation: curve (a) is drawn for
$W\left( a_{m}\right) =0.01\%$;
curve (b) for $W\left( a_{m}\right) =0.1\%$; curve (c) for $%
W\left( a_{m}\right) =1\%$.}
\label{figure4}
\end{figure}

\begin{figure}[tbp]
\epsfbox{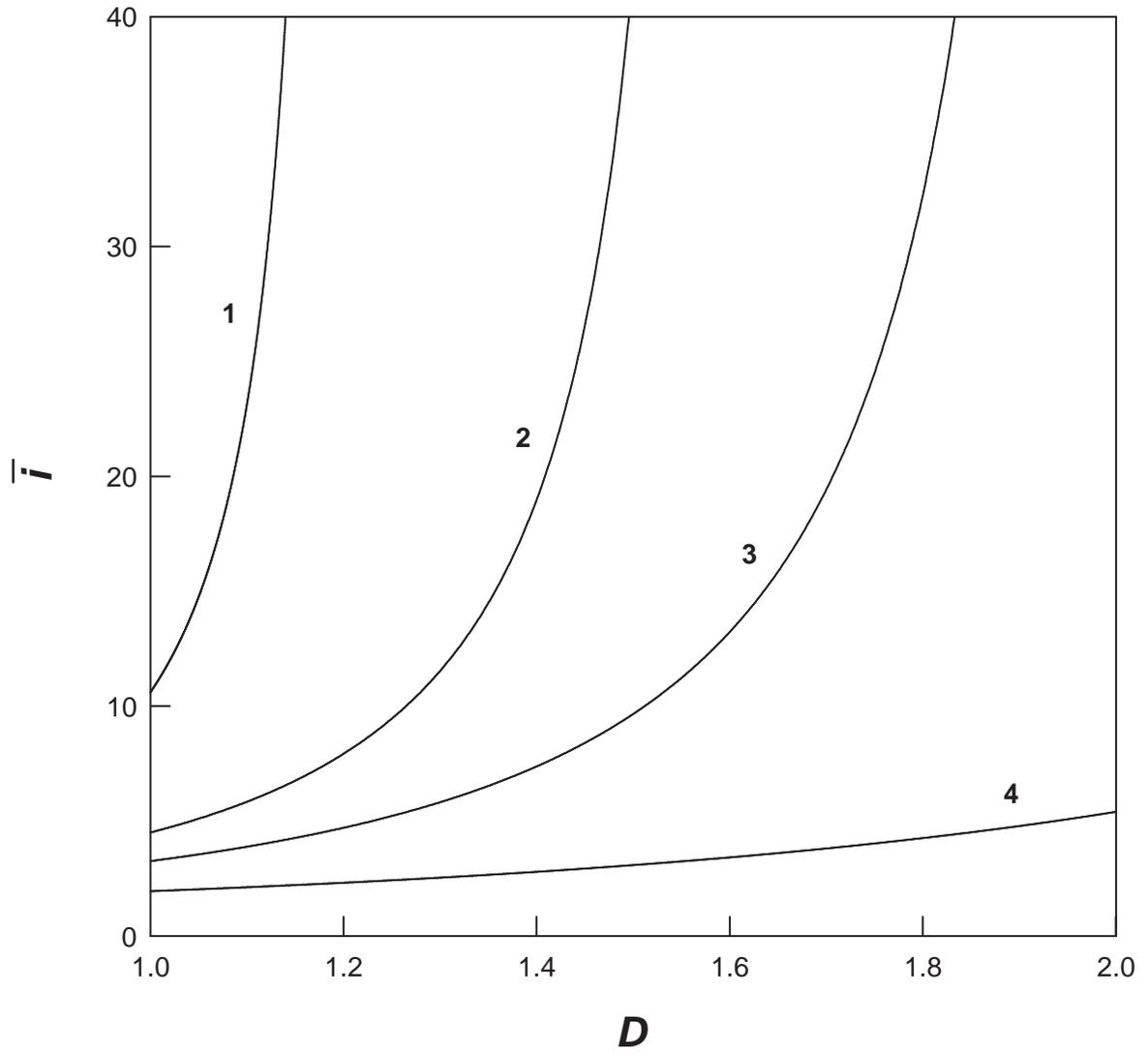}
\caption{Influence of the fractal dimension of the cluster boundary on the
mean critical current. Curve (1) corresponds to the case of $g=-0.4$;
curve (2) of $g=-0.2$; curve (3) of $g=0$; curve (4) of $g=1$.}
\label{figure5}
\end{figure}

\begin{figure}[tbp]
\epsfbox{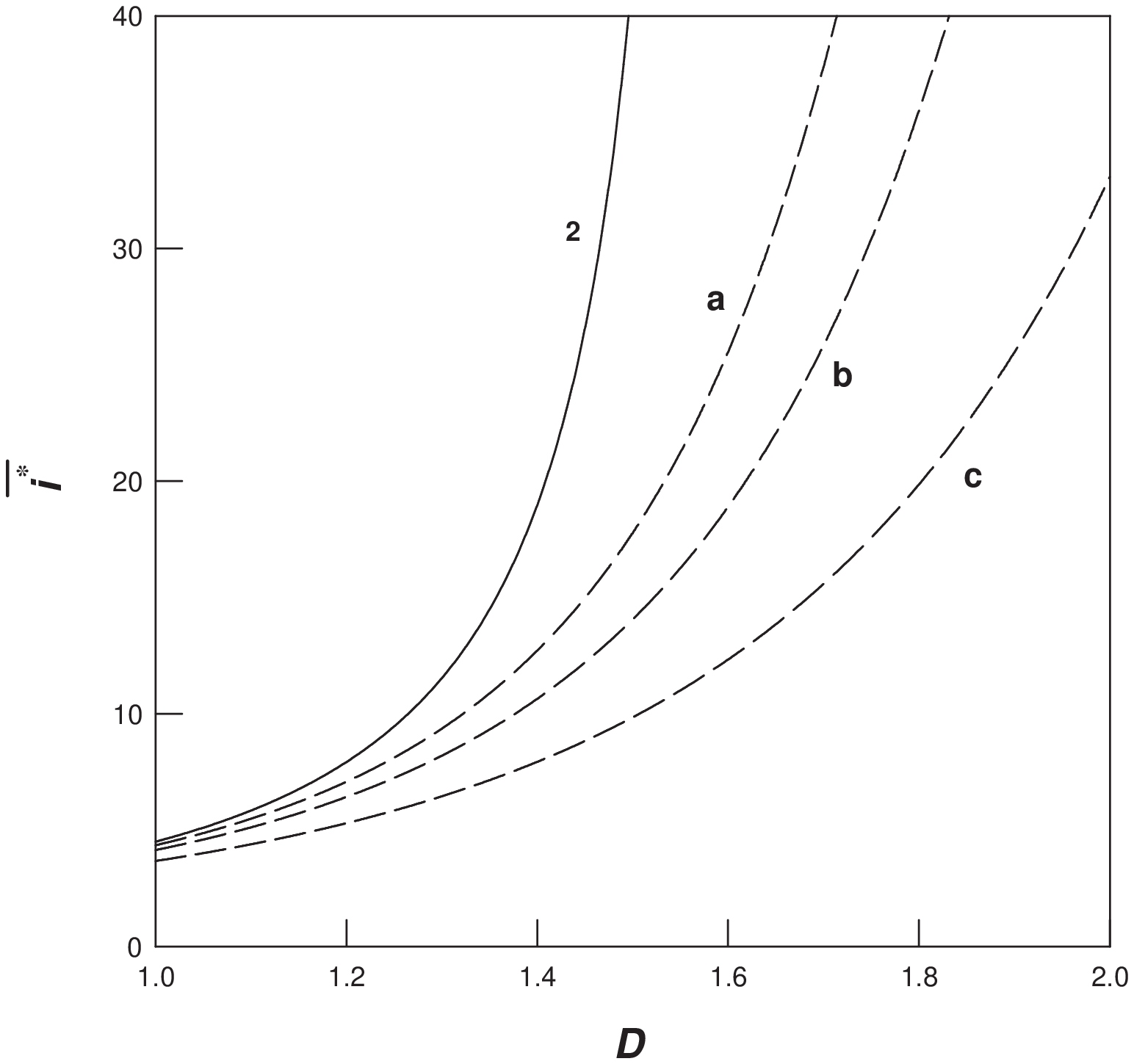} \caption{Mean critical current graphs for
truncated distribution at $g=-0.2$ with the different degree of
truncation: curve (2) is drawn for $W\left( a_{m}\right) =0$;
curve (a) for $W\left( a_{m}\right) =0.01\%$; curve (b) for
$W\left( a_{m}\right) =0.1\%$; curve (c) for
$W\left(a_{m}\right)=1\%$. The dotted lines are given for
truncated distribution, the solid line for initial distribution.}
\label{figure6}
\end{figure}

For the truncated distribution of Eq.~(\ref{truni8}) the possible
values of depinning currents are bounded from above by the
quantity $i_{m}$, therefore the mean critical current as well as
the variance do not diverge any more. Both of these
characteristics are finite for any fractal dimensions, including
the case of maximum fractality ($D=2$). The corresponding graphs
for standard deviation are presented in Fig.~\ref{figure4}. All
the curves are calculated at $g=-0.2$ (as for the main curve (2)
of the mean critical current in Fig.~\ref{figure5}). In this case
the variance for initial distribution of the critical currents is
infinite, so no graph for $W\left( a_{m}\right) =0$ is shown in
Fig.~\ref{figure4} at all. The dependence of the mean critical
current on the fractal dimension in the case of truncated
distribution is demonstrated in Fig.~\ref{figure6}. The
corresponding graphs in Figs.~\ref{figure4} and \ref{figure6} are
drawn for the same values of truncation degree. The truncation
degree is related to the least possible area of the cluster by the
equation (\ref{degr7}). The values of $W\left( a_{m}\right) $ and
$a_{m}$ involved in Figs.~\ref{figure4} and \ref{figure6} are
presented in the Table~\ref{table1}.

The degree of truncation gives the probability measure of the number of
normal phase clusters that have the area smaller than the lower bound $a_{m}$
in the initial distribution. If the magnitude of $W\left( a_{m}\right) $ is
sufficiently small (no more than several percent), then the procedure of
truncation scarcely affects the initial shape of the distribution and still
enables the contribution from the clusters of zero area (therefore, of
infinitely high depinning current) to be eliminated. It is interesting to
note that the very similar situation occurs in analyzing the statistics of
the areas of fractal islands in the ocean \cite{mandelbrot}. The island
areas obey the Pareto distribution, which also has the hyperbolic
singularity at zero point that causes certain of its moments to diverge. For
exponential distribution of the cluster areas, which is valid in the
above-mentioned case of YBCO films \cite{prb}, the probability to find the
cluster of area smaller than $a_{m}$ in the sampling is equal to $\Pr
\left\{ \forall a<a_{m}\right\} =2.7\%$ only. In principle, the truncation
procedure could be made here, too, but there is no need for that, because
the contribution of infinitesimally small clusters is finite at $g=0$ (and
equal to zero at $g>0$).

Referring to Figs.~\ref{figure4} and \ref{figure6} it can be seen that the
dependencies of the standard deviation and the mean critical current on the
fractal dimension become smoother and smoother with increase in the degree
of truncation. At the same time, the inherent tendency of initial
distribution is still retained: as the fractal dimension increases, the
critical current distribution broadens out (the variance grows, see Fig.~\ref
{figure4}), moving towards higher currents (the mean critical current grows,
see Fig.~\ref{figure6}). As may be seen from Fig.~\ref{figure2}, this trend
is further enhanced with decreasing $g$-parameter. The most current-carrying
capability of a superconductor should be achieved when the clusters of small
size, which have the highest currents of depinning, contribute maximally to
the overall distribution of the critical currents. Such a situation takes
place just in the region of giant variance of critical currents. So far, the
least magnitude of $g$-parameter (equal to zero) has been realized in YBCO
composites containing normal phase clusters of fractal dimension $D=1.44$
\cite{pla}. The critical currents of superconducting films with such
clusters are higher than usual \cite{tpl,pjtf99,ftt99}. It would thus be
expected to further improve the current-carrying capability in
superconductors containing normal phase clusters, which will be
characterized by area distribution with negative magnitudes of $g$%
-parameter, especially at the most values of fractal dimensions.

Thus, it has been revealed that the distribution of the critical currents in
superconductor with fractal clusters of a normal phase has the anomalous
statistical properties, implying that its variance and expectation diverges
in the certain range of fractal dimensions. It may be expected that the most
current-carrying capability of a superconductor can be achieved by
optimization of the cluster area distribution that involves reducing
g-parameter with concurrent increasing the fractal dimension.

\section{Acknowledgements}

This work is supported by the Russian Foundation for Basic Researches (Grant
No 02-02-17667).

\begin{table}[tbp] \centering%
%
\caption{Degree of truncation of the cluster area distribution and
corresponding least possible area of the normal phase cluster at g
= - 0.2\label{table1}}
\begin{tabular}{ccc}
Truncation degree $W\left( a_{m}\right) $ & Minimum area $a_{m}$ & Markers
in Figs. 4, 6 \\ \hline
$0.01\%$ & $1.144\times 10^{-5}$ & {\sf a} \\
$0.1\%$ & $2.034\times 10^{-4}$ & {\sf b} \\
$1\%$ & $3.622\times 10^{-3}$ & {\sf c}
\end{tabular}
\end{table}%
%

\newpage

\newpage

\newpage

\newpage

\newpage

\newpage

\begin{references}
\bibitem{surdeanu}  R. Surdeanu, R. J. Wijngaarden, B. Dam, J. Rector, R.
Griessen, C. Rossel, Z. F. Ren, and J. H. Wang, Phys. Rev. B {\bf 58}, 12467
(1998); C. J. Olson, C. Reichhardt, and F. Nori, Phys. Rev. Lett. {\bf 80},
2197 (1998).

\bibitem{pla}  Yu. I. Kuzmin, Phys. Lett. A {\bf 267}, 66 (2000).

\bibitem{prb}  Yu. I. Kuzmin, Phys. Rev. B {\bf 64}, 094519 (2001).

\bibitem{bazil}  M. Baziljevich, A. V. Bobyl, H. Bratsberg, R. Deltour, M.
E. Gaevski, Yu. M. Galperin, V. Gasumyants, T. H. Johansen, I. A. Khrebtov,
V. N. Leonov, D. V. Shantsev, and R. A. Suris, J. Phys. (Paris) IV {\bf 6},
C3-259 (1996); M. Prester, Supercond. Sci. Technol. {\bf 11}, 333 (1998); M.
Prester, Phys. Rev. B {\bf 60}, 3100 (1999); B. Dam, J. M. Huijbregtse, and
J. H. Rector, Phys.Rev. B {\bf 65}, 064528 (2002).

\bibitem{mezzetti}  E. Mezzetti, R. Gerbaldo, G. Ghigo, L. Gozzelino, B.
Minetti, C. Camerlingo, A. Monaco, G. Cuttone, and A. Rovelli, Phys.Rev. B
{\bf 60}, 7623 (1999).

\bibitem{tpl}  Yu. I. Kuzmin, Tech. Phys. Lett. {\bf 26}, 791 (2000).

\bibitem{pla2}  Yu. I. Kuzmin, Phys. Lett. A {\bf 281}, 39 (2001).

\bibitem{pss}  Yu. I. Kuzmin, Phys. Solid State {\bf 43}, 1199 (2001).

\bibitem{pjtf99}  Yu. I. Kuzmin and I. V. Plechakov, Tech. Phys. Lett. {\bf %
25}, 475 (1999).

\bibitem{ftt99}  Yu. I. Kuzmin, I. V. Pleshakov, and S. V. Razumov, Phys.
Solid State {\bf 41}, 1594 (1999).

\bibitem{duran}  C. A. Duran, P. L. Gammel, R. Wolfe, V. J. Fratello, D. J.
Bishop, J. P.Rice, and D. M. Ginsberg, Nature (London) {\bf 357}, 474
(1992); C. A. Duran, P. L. Gammel, D. J. Bishop, J. P. Rice, and D. M.
Ginsberg, Phys. Rev. Lett. {\bf 74}, 3712 (1995); U. Welp, T. Gardiner, D.
O. Gunter, B. W. Veal, G. W. Crabtree, V. K. Vlasko-Vlasov, and V. I.
Nikitenko, Phys. Rev. Lett. {\bf 74}, 3713 (1995); R. J. Wijngaarden, R.
Griessen, J. Fendrich, and W.-K. Kwok, Phys. Rev. B {\bf 55}, 3268 (1997).

\bibitem{mandelbrot}  B. B. Mandelbrot, {\it The Fractal Geometry of Nature}
(Freeman, San Francisco, 1982).

\bibitem{mandelbrot2}  B. B. Mandelbrot, {\it Fractals: Form, Chance, and
Dimension} (Freeman, San Francisco, 1977).

\bibitem{hudson}  D. Hudson, {\it Statistics} (CERN, Geneva, 1964).
\end{references}
\end{document}